\documentclass[
    ,final            
  ]
  {aipproc}

\layoutstyle{6x9}

\usepackage{psfig}
\usepackage{amsmath,bbm,amssymb}

\def\gtap{\mathrel{\hbox{\rlap{\lower.55ex \hbox {$\sim$}}
                   \kern-.3em \raise.4ex \hbox{$>$}}}}
\def\ltap{\mathrel{\hbox{\rlap{\lower.55ex \hbox {$\sim$}}
                 \kern-.3em \raise.4ex \hbox{$<$}}}}

\begin{document}

\title{X-ray sources in globular clusters}


\author{Frank Verbunt}{
  address={Astronomical Institute, Postbox 80.000, 3508 TA Utrecht,
the Netherlands}
}

\begin{abstract}
Observations with BeppoSAX, RXTE and Chandra suggest that many of the
bright X-ray sources in globular clusters have ultrashort binary
periods.  This is remarkable as such systems are not easily
formed. With accurate optical astrometry of HST images, the large
numbers of low-luminosity X-ray sources discovered with Chandra can be
classified as quiescent low-mass X-ray binaries, pulsars, cataclysmic
variables, and magnetically active binaries. The number of cataclysmic
variables is found to scale with the number of close stellar
encounters.
\end{abstract}

\maketitle


\section{Introduction}

The known populations of X-ray sources in globular clusters have
grown with the increasing sensitivity of X-ray detectors, from
bright low-mass X-ray binaries discovered in the 1970s (UHURU, OSO-7,
Ariel-V), to faint low-mass X-ray binaries and cataclysmic variables
in the 1980s (Einstein, HEAO-2), recycled radio pulsars in the 1990s
(ASCA, ROSAT), and magnetically active close binaries in the 2000s
(Chandra). In this review I summarize the new knowledge gained in the
last years about the bright X-ray sources, and discuss in particular
the possibility that many of these have ultrashort orbital periods
(Sect.\,2).  Then I explain how various new technical possibilities of
Chandra, XMM and HST expand our knowledge of the faint sources
(Sect.\, 3). Some considerations on formation rates end the paper.

\section{The bright sources: neutron stars accreting at high rates}

We now know 13 bright X-ray sources 
($L_{0.5-2.5\,{\mathrm keV}}>10^{35.5}$\,erg\,s$^{-1}$,
say) in 12 globular clusters of our Galaxy (Table\,\ref{ta:bright}).
Most were discovered in the 1970s with the first X-ray satellites;
relatively recent additions are the source in Terzan\,6, discovered in
the ROSAT All Sky Survey, and the two sources into which the bright
X-ray emission from NGC\,7078 -- known since the 1970s -- was resolved
by Chandra.  X-ray bursts were detected from some sources soon after
discovery, and some sources (Terzan\,1 and Terzan\,5) were initially
detected only during bursts.  It required the large temporal and
spatial coverage by the BeppoSAX Wide Field Cameras and the RXTE All
Sky Monitor to find that all but one of the bright sources in globular
clusters are X-ray bursters, and thus are neutron stars accreting from
a companion.  The 13th source, in NGC7078, may be a neutron star as
well.

\begin{table}
\centerline{
\begin{tabular}{llllllll}
\phantom{xxxxxxxxx}cluster    & disc         & burst    &
 $M_{\lambda}$   &   $P_b$ & TOXB \\   
\phantom{xxxxxxxxx}NGC\,$1851$ & 1975 & 1976 & 5.6B & & \phantom{T}UUU\\
\phantom{xxxxxxxxx}NGC\,$6440$ & 1975 & 1999 & 3.7B & & T\phantom{}$-$N$-$& \\
\phantom{xxxxxxxxx}NGC\,$6441$ & 1974 & 1987 & 2.4B & 5.7hr &  \phantom{T}$-$NN \\
\phantom{xxxxxxxxx}NGC\,$6624$ & 1974 & 1976 & 3.0B & 11.4m & \phantom{T}UUU \\
\phantom{xxxxxxxxx}NGC\,$6652$ & 1985 & 1998 & 5.6B &  & \phantom{T}UUU\\
\phantom{xxxxxxxxx}NGC\,$6712$ & 1976 & 1980 & 4.5B & 20.6m$^b$ & \phantom{T}UUU \\
\phantom{xxxxxxxxx}NGC\,$7078$-1$^a$ & 2001 & & 0.7B & 17.1hr & \phantom{T}$-$$-$$-$\\
\phantom{xxxxxxxxx}NGC\,$7078$-2$^a$ & 2001 & 1990 & 3.1U & & \phantom{T}$-$$-$U \\
\phantom{xxxxxxxxx}Terzan\,1   & 1981 & 1981  &    & & T\phantom{}$-$$-$$-$   \\
\phantom{xxxxxxxxx}Terzan\,2   & 1977 & 1977 & & & \phantom{T}$-$NU\\
\phantom{xxxxxxxxx}Terzan\,5   & 1981 & 1981 &1.7J & & T\phantom{}$-$U$-$ \\
\phantom{xxxxxxxxx}Terzan\,6   & 1991 & 2003& & 12.36h & T\phantom{}$-$N$-$ \\
\phantom{xxxxxxxxx}Liller\,1   & 1976 & 1978 & & & T\phantom{}$-$$-$$-$ \\
\end{tabular}
\caption{\it List of the thirteen bright X-ray sources in globular
clusters with the year of discovery, year of the first observation of a
burst, absolute magnitude in $B$ (or $U,J$, as indicated), orbital
period, and indications whether the source is a transient (T),
and whether the source has properties associated with an ultrashort period
(U) or with a normal period (N) in its optical magnitude (O), X-ray
spectrum (X), or burst maximum (B). See text for explanation. Detailed
references are given in Verbunt \&\ Lewin (2004).
$^a$Luminous X-ray emission from NGC\,7078 was already found in 1974.
$^b$or its alias at 13.2\,m.
\label{ta:bright}}
}
\end{table}
\nocite{vl04}

Remarkably, of the five orbital periods currently known, two are ultrashort,
at periods of 11.4\,m and 20.6\,m (or its alias 13.2\,m). I return to this
below. Five of the 13 sources are transients, in the sense that their
luminosity at times has dropped below $\sim10^{34}$\,erg\,s$^{-1}$.

\begin{figure}
\centerline{\psfig{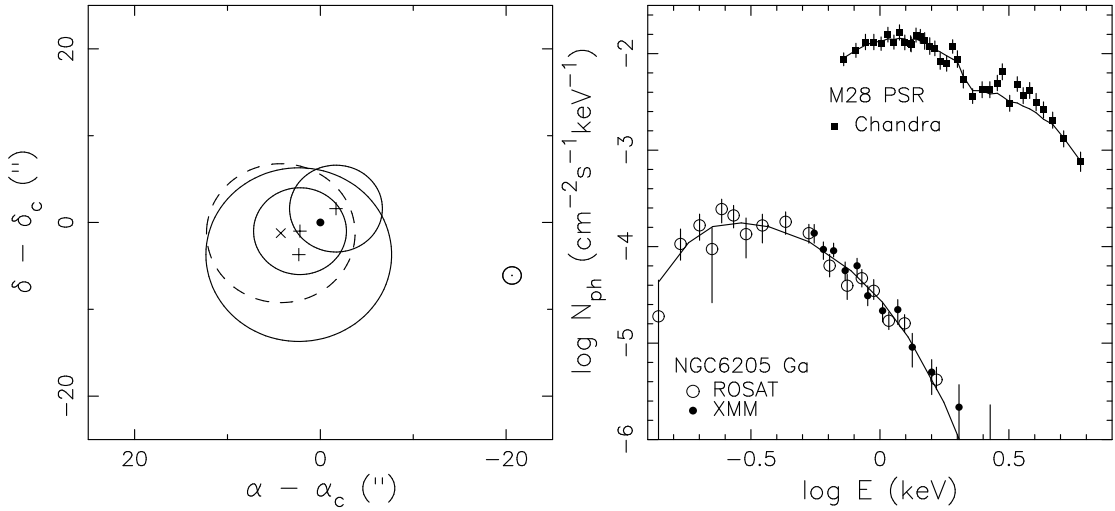}}

\caption{\it Left: Accurate positions for the bright source
in Terzan\,1 were determined with EXOSAT ($\times$ with dashed error
circle, from Parmar et al.\ 1989) and with three independent ROSAT
observations (+ with solid error circles, redetermined by the author
after Johnston et al.\ 1995). Coordinates are with respect to the
cluster center. The bright source had switched off
in 1999 (Guainazzi et al.\ 1999). The position of
the one source detectable in a short Chandra observation, indicated as
the small error circle on the right, is {\it not} compatible with
the position of the previously bright source (Wijnands et al.\ 2002).
Right: Comparison of the X-ray spectrum of a quiescent low-mass
X-ray binary (NGC6205 Ga, below left) and a radio pulsar (in M28,
up right). Note the sensitivity of ROSAT to very soft photons.
Data from ROSAT and XMM from Gendre et al.\ (2003); from 
Chandra from Becker et al.\ (2003).
\label{verbfa}}
\end{figure}
\nocite{psg89}\nocite{jvh95}\nocite{gpo99}\nocite{whg02}\nocite{gbw03b}
\nocite{bsp+03}

Recently, two remarkable illustrations of the danger of
identifications without accurate positions have been found. The first
is the discovery that the bright ultraviolet source in NGC\,6652, with
a period of 43.6\,m (Deutsch et al.\ 2000), is {\it not} the optical
counterpart of the bright X-ray source in this cluster, but of a
second, relatively faint X-ray source, presumably a faint low-mass
X-ray binary (Heinke et al.\ 2001).  The second is the discovery that
the single faint source significantly detected in a Chandra
observation of Terzan\,1 is {\it not} the quiescent counterpart of the
transient in this cluster, but a different source (see
Figure\,\ref{verbfa}).  A warning that the optical centers of globular
clusters are not always as accurate as advertized is provided by the
study of Terzan\,6: the hitherto assumed optical center was probably
offset from the actual center by the influence of a bright star
outside the cluster. A re-determination of the cluster center, in
which each star gets the same weight irrespective of its brightness,
places the center very close to the accurately determined position of
the X-ray source (in 't Zand et al.\ 2003).
\nocite{dma00}\nocite{heg01}\nocite{zhm+03}

With three ROSAT observations of NGC\,6624, van der Klis et al.\
(1993) showed that the 11.4\,m period of the X-ray source in this
cluster decreases, in contradiction of the model in which the donor
delivering matter to the neutron star is a white dwarf.  The negative
period derivative is confirmed by Chou \&\ Grindlay (2001). Some
hesitance in accepting this derivative as intrinsic may remain, since
the differences between observed periods as a function of time and a
quadratic fit look irregular. Also, it is possible that acceleration
of the binary in the cluster potential affects its observed period
derivative: the optical counterpart of the X-ray source is very close
to the cluster center ($0.6\pm0.3''$, King et al.\ 1993).
Nonetheless, one must now seriously consider the possibility that the
11.4\,m period is decreasing rather than increasing.
\nocite{khv+93}\nocite{cg01}\nocite{ksa+93}

Are there models which can explain this? To discuss this, we note
first that the formation of the bright X-ray binaries in globular
clusters is thought to follow two main paths: either a tidal capture
of a closely passing star by a single neutron star, or an exchange
encounter in which a neutron star encounters a binary and swaps place
with one binary star. If a low-mass main-sequence star is captured,
and starts transferring mass to the neutron star, the orbit will
decrease until the donor star becomes degenerate, after which the
orbit expands again. The minimum period reached this way is near
70-80\,m, and shorter periods cannot be explained this way.  If a
more massive main-sequence star is captured, and evolves into
a giant before mass transfer starts, the mass transfer is unstable and
a spiral-in of the neutron star into the envelope of the giant may
bring it into a close orbit with the core of the giant. This must have
happened in the past, when stars with a mass sufficiently high to induce
unstable mass transfer still
existed in globular clusters: it may take several billion years before
the core of the giant, cooled into a white dwarf, comes into contact
under the influence of gravitational radiation
and starts mass transfer. Such ultrashort period binaries can
therefore be observed today (Rasio et al.\ 2000).  If a neutron star
collides directly with a giant, the envelope will be expelled, and the
core may enter an orbit around the neutron star. A white-dwarf donor
of mass $M_c$ fills its Roche-lobe at an orbital period of roughly
$P_b\sim 1^{\rm m} (M_\odot/M_c)$; thus ultrashort periods are
possible, but the period must lengthen with time as the mass of the
white dwarf decreases (Verbunt 1987).  Tutukov et al.\ (1985) note
that ultrashort periods can also be reached in binaries in which a
star starts to transfer mass to its companion after it has already
evolved somewhat, and depleted its core from hydrogen. The donor then
becomes degenerate at smaller mass, and shorter orbital
period. Tutukov et al.\ do a limited number of calculations and find a
minimum period of about 20\,m; Podsiadlovski et al.\ (2002) show that
periods as short as 5\,m can be reached this way.  This scenario may
be called 'magnetic capture'.
\nocite{rpr00}\nocite{ver87}\nocite{tfey85}\nocite{prp02}

Van der Sluys et al.\ (these proceedings, see also Van der Sluys et
al.\ 2004) note that none of the relevant computations of magnetic
capture done by Pylyser \&\ Savonije (1988) reach such extremely short
periods, and investigate in some detail which initial conditions
(initial binary period and donor mass) lead to ultrashort orbital
periods {\it within a Hubble time}. They find that only binaries in a
very small range of initial orbital periods and in a small range of
donor masses evolve to ultrashort periods within the Hubble time, and
that each of these binaries spends only a small fraction of its
evolution at ultrashort periods. They conclude that only an extremely
small fraction ($<0.001$) of a population of X-ray binaries at any
time can be expected to be at ultrashort periods.  Whereas the
observation of one such system amongst the thirteen globular cluster
sources could be explained by magnetic capture, the observation of two
or more demands another explanation.  \nocite{ps88}\nocite{svp04}

\begin{figure}
\centerline{\psfig{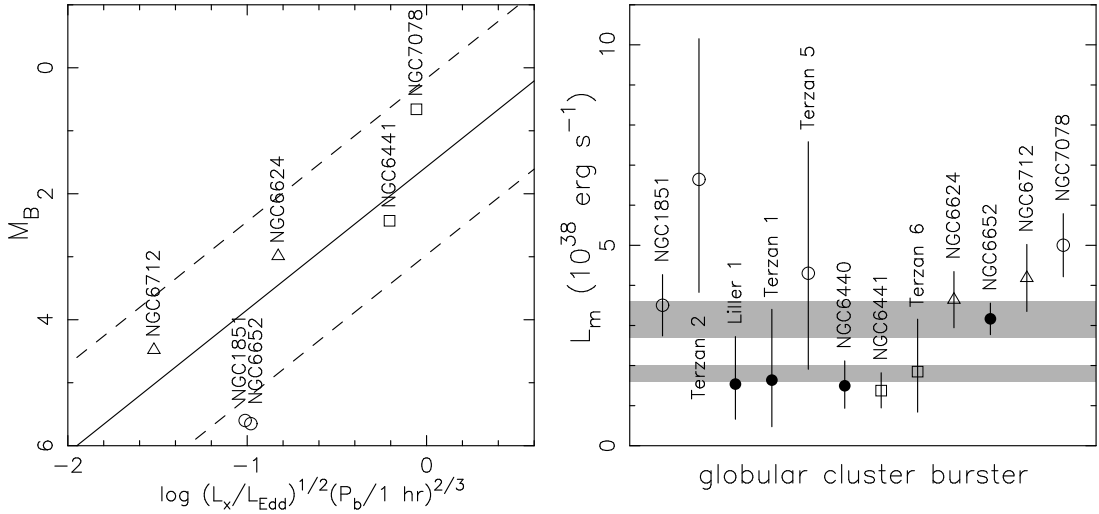}}

\caption{\it Left: Absolute B magnitude of optical 
counterparts to X-ray sources in globular clusters 
as a function of the X-ray luminosity
(expressed in Eddington luminosity $2\times10^{38}$erg\,s$^{-1}$)
and orbital period (in hours). Data from Deutsch et al.\ (2000)
and Heinke et al.\ (2001), see Table\,\ref{ta:bright}. The solid and 
dashed lines give the
average relation and spread around the average for X-ray binaries
in the galactic disk, as derived by Van Paradijs \&\ McClintock (1994).
Right: maximum observed luminosities $L_m$ during X-ray bursts, compared with
the Eddington limits for hydrogen-rich ($X=0.73$) and hydrogen-poor ($X=0$)
matter (lower and upper gray band, respectively). After Kuulkers et 
al.\ (2003). In both figures:
$\Box$ normal period, $\triangle$ ultrashort period, $O$ period
not known (set at 1\,hr for the left plot). A filled symbol indicates that
the peak flux was determined for an ordinary burst, and may be
lower than the Eddington limit.
\label{verbfb}}
\end{figure}
\nocite{dma00}\nocite{heg01}\nocite{vpm94}\nocite{khz+03}

In this context it is worthy of note that there are good indications
that many of the bright X-ray sources in
globular clusters have ultrashort periods, as first pointed out by
Deutsch et al.\ (1996). As seen in Table\,\ref{ta:bright}, two of five
orbital periods known are ultrashort.  Indirect evidence for
ultrashort periods is provided by three observations.  First, it is
known that the optical light of bright low-mass X-ray binaries is
dominated by reprocessing of X-rays that impinge on the accretion
disk.  Thus, if the disk is small and/or if the X-ray luminosity is
small, one expects a low optical luminosity. Van Paradijs \&\
McClintock (1994) derive that the expected absolute visual magnitude
scales with disk size (via orbital period $P_b$) and X-ray luminosity
$L_x$ roughly as $M_V\propto
{P_b}^{2/3}{L_x}^{1/2}$. Figure\,\ref{verbfb} shows that the
transition between systems with normal periods (i.e.\ periods above
80\,m) and those with ultrashort periods lies near $M_B=3$. The B
magnitudes of the sources in NGC\,1851 and NGC\,6652 thus suggest
ultrashort periods.  \nocite{damd96}\nocite{vpm94}

\begin{figure}
\centerline{\psfig{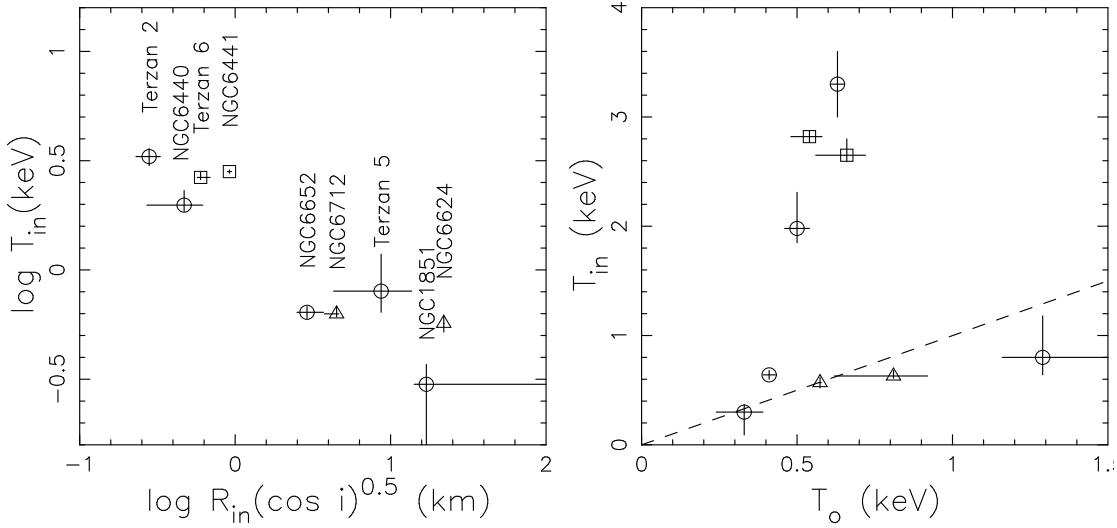} {\hfil}}

\caption{\it Fitted values of model parameters of X-ray spectra
of X-ray sources in globular clusters. Shown are the blackbody temperatures
$T_{in}$ of the photons leaving the inner disk as a function of the
projected inner disk radius $R_{in}\sqrt{\cos(i)}$ (left) and of the blackbody
temperatures $T_o$ of photons entering the Comptonization region (right).
After Sidoli et al.\ (2001); NGC\,6652 added from Parmar et al.\ 2001,
Terzan\,5 added from Heinke et al.\ (2003). Symbols as in Fig.\,\ref{verbfb}.
\label{verbfc}
}
\end{figure}
\nocite{spo+01}\nocite{pos+01}\nocite{heg+03}

The second indication is the maximum luminosity reached in a burst.
Kuulkers et al.\ (2003) have analyzed an extensive data set of X-ray
bursts of sources in globular clusters, and note that the maximum
luminosity reached in radius expansion bursts, when the source is
assumed to reach the Eddington limit, corresponds to the Eddington
limit for hydrogen-poor material in the two systems with ultrashort
periods, and to that for hydrogen-rich material in the source with a
normal period in NGC\,6641. The maximum luminosities in the other
cluster sources then suggest ultrashort periods in NGC\,1851,
NGC\,6652, NGC\,7078-2, and Terzan\,2 (Fig.\,\ref{verbfb}).
\nocite{khz+03}

A third indication comes from the X-ray spectra. A model which is
often used -- but debated as to its physical correctness -- is that in
which a sum of blackbody spectra from annuli of the accretion disk is
combined with a Comptonized spectrum. The parameters are the inner
radius and inner temperature of the disk, and the temperature of the
photons entering the Comptonized region, characterized by an optical
depth and electron temperature.  The best-fit parameters appear to
separate the ultrashort systems in NGC\,6624 and NGC\,6712, for which
the inner radius is compatible with a neutron star radius and for
which the photons entering the Comptonization region have a
temperature compatible with those leaving the disk, from the systems
with normal periods in NGC\,6441 and Terzan\,6, for which the inner
disk radii are too small, and the photons entering the Comptonization
region too soft, for the model to be physically viable (Sidoli et al.\
2001).  This suggests that the sources in NGC\,1851, NGC\,6652 and
Terzan\,5 also have an ultrashort period, and those in NGC\,6440 and
Terzan\,2 normal periods (Sidoli et al.\ 2001, Parmar et al.\ 2001,
Heinke et al.\ 2003).  \nocite{spo+01}\nocite{pos+01}\nocite{heg+03}

Collating these three observational indicators, Verbunt \&\ Lewin
(2004) find that the sources in NGC\,1851 and NGC\,6652 probably, and
those in NGC\,7078-2 and Terzan\,2 possibly, have ultrashort periods 
(Table\,\ref{ta:bright}).
The source in NGC\,6440 possibly has a normal period. The indications
are contradictory in Terzan\,2 and absent in NGC\,7078-1 and
Liller\,1.  Thus, certainly 2, probably 4, and possibly 6 or 8 of the
thirteen bright X-ray sources in globular clusters have ultrashort
periods!  \nocite{vl04}

\begin{figure}
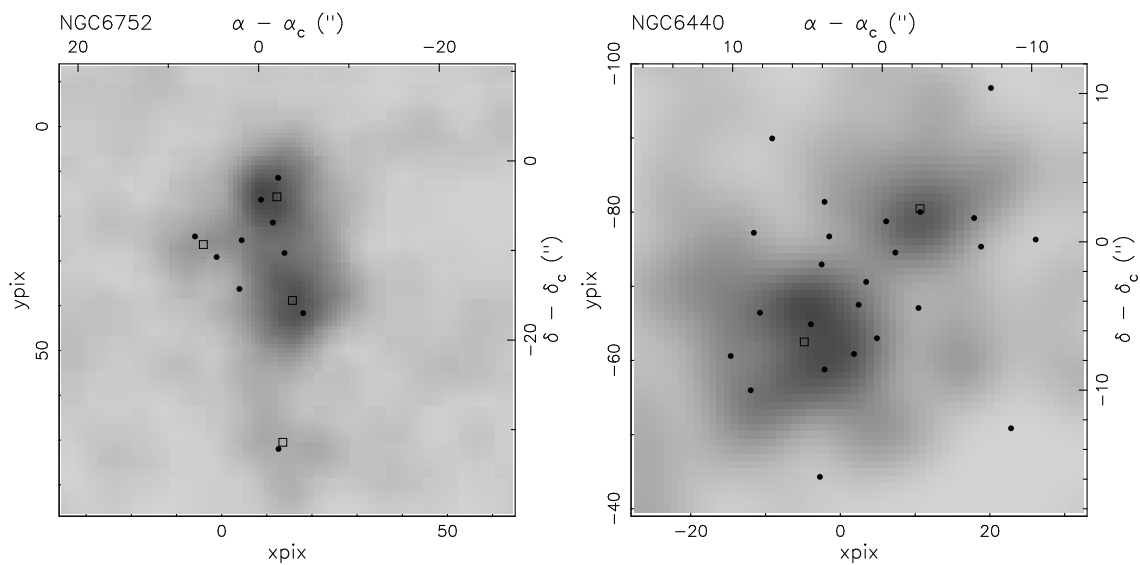

\centerline{
\parbox[b]{7.5cm}{\psfig{figure=verbfda.ps,width=7.5cm,clip=t}}
\parbox[b]{7.5cm}{\psfig{figure=verbfdb.ps,width=7.5cm,clip=t}}
}
\caption{\it Smoothed grey scale images of NGC\,6752 and NGC\,6440 as obtained
with ROSAT, with the positions of the sources derived from the ROSAT
data ($\Box$) and the much more numerous positions of the sources detected
with Chandra ($\bullet$), illustrating the dramatic increase in resolution
and sensitivity. The uncertain direction (`boresight') 
of the ROSAT observation allows
a shift between the ROSAT and Chandra image of NGC\,6440 by 5$''$.
ROSAT data from Verbunt \&\ Johnston (2000) and Verbunt et al.\ (2000);
Chandra positions from Pooley et al.\ (2002a,b).
\label{verbfd}}
\end{figure}
\nocite{vj00}\nocite{vkzh00}\nocite{plh+02}\nocite{plv+02}

\section{Various techniques to study the dim sources}
 
CCDs have a better {\bf spectral resolution} than proportional
counters, and as a result the spectral energy distribution of X-ray
sources is better determined from observations with Chandra and XMM
than with ROSAT (Fig.\,\ref{verbfa}). If a source has a soft spectrum, with a
characteristic blackbody temperature of $\ltap0.3$\,keV say, and an
X-ray luminosity $\gtap10^{32}$erg\,s$^{-1}$, then it is almost
certain to be a neutron star accreting from a companion. With this
method an increasing number of faint accreting neutron stars has been
discovered in globular clusters, showing that the number of such
sources is more than ten times higher than the number of bright
sources (Pooley et al.\ 2003). In general, accreting neutron stars are
close to the cluster center, within about 2 core radii, but of order
20\%\ are further from the core. For example, the bright source in
NGC\,6652 is about six core radii from the center of that cluster
(Heinke et al.\ 2001).  With the accuracy of Chandra positions, the
error in the position of a source with respect to the cluster center
is now dominated by the uncertainty in the cluster center.
\nocite{pla+03}\nocite{heg01}

The Chandra telescope has a much better {\bf spatial resolution} than
ROSAT, and as a result many more faint
sources have been discovered in clusters (Figure\,\ref{verbfd}).  The
improved accuracy of the source positions allows a fairly safe
identification of X-ray sources with accurately localized
radio pulsars on the basis of
positional coincidence alone. A dozen pulsars has thus been detected
in 47\,Tuc without ambiguity; two more sources are blends of two
pulsars each. A single pulsar has been detected in each of the
clusters NGC\,6397, NGC\,6752 and M\,28 (Grindlay et al.\ 2002,
D'Amico et al.\ 2002, Becker et al.\ 2003). The latter source, already
detected with ASCA and ROSAT (Saito et al.\ 1997, Verbunt 2001), is
very bright, and shows that the total X-ray luminosity in the
0.5-2.5\,keV range scales directly with the spindown luminosity, as is
the case for pulsars (both young and recycled) in the disk of the
galaxy (Verbunt et al.\ 1996).
\nocite{gch+02}\nocite{dpf+02}\nocite{bsp+03}\nocite{skk+97}\nocite{ver01}
\nocite{vkb+96}

The {\bf positional accuracy} of Chandra and of the Hubble Space Telescope
allows much more secure identification of optical counterparts to the X-ray
sources than earlier ROSAT data. To take full advantage of this opportunity,
one must align the coordinates from Chandra and HST.
From the absolute position of a Chandra source (with accuracy
0.6$''$, say) and of the HST objects (with accuracy of 1$''$, say;
but occasionally as big a 3$''$) we must look for counterparts
in the HST image in a circle with a radius of $\sqrt{0.6^2+1^2}=1.2''$,
or occasionally even larger.
On the other hand, once we have securely identified at least one
Chandra source with an HST object, we can align the coordinate frames
with this match, and the error circles for the
remaining sources are much reduced: for sufficiently bright
X-ray sources the relative positions are accurate to $\ltap0.1''$,
the relative optical positions are accurate to  $\ltap0.1''$,
and thus the radius of the search circle now is $\sqrt{0.14^2+{\sigma_m}^2}''$,
where $\sigma_m$ is the accuracy of the match.

In the case of a rare source type, like pulsars, identifications are
secure even with the larger error circles of the absolute astrometry;
to be useful for optical identifications, the pulsar radio position must
be tied to the optical coordinate frame. A beautiful example of this
is the identification of the pulsar in NGC6397 with a Chandra source,
on the basis of the binary period, detected both in the radio and in the
optical (Ferraro et al.\ 2001). If such a direct match is not
available, one may improve the HST absolute positions by
connecting its coordinate system to absolute astrometry. For this
purpose, the USNO CCD Astrograph Catalog (UCAC: Zacharias et al.\ 2000)
turns out to be excellent, as shown by Bassa et al.\ (2004) for 
NGC\,6121. Bassa et al.\ first identify 91
stars in an ESO 2.2m Wide Field Imager observation with stars in 
UCAC to determine the coordinates of the ESO WFI image, and
then identify about 200 stars in the HST WFPC2 image with stars on the
ESO WFI image to determine the coordinates of the HST frame
with a 1\,$\sigma$ accuracy of $0.12''$ both in right ascension and
in declination. This allows identification of three optical objects
with Chandra sources, which are then used to align the Chandra frame
to the HST frame, with an accuracy of about 0.14$''$.
\nocite{fpas01}\nocite{zuz+00}\nocite{bph+04}

\begin{figure}
\centerline{\psfig{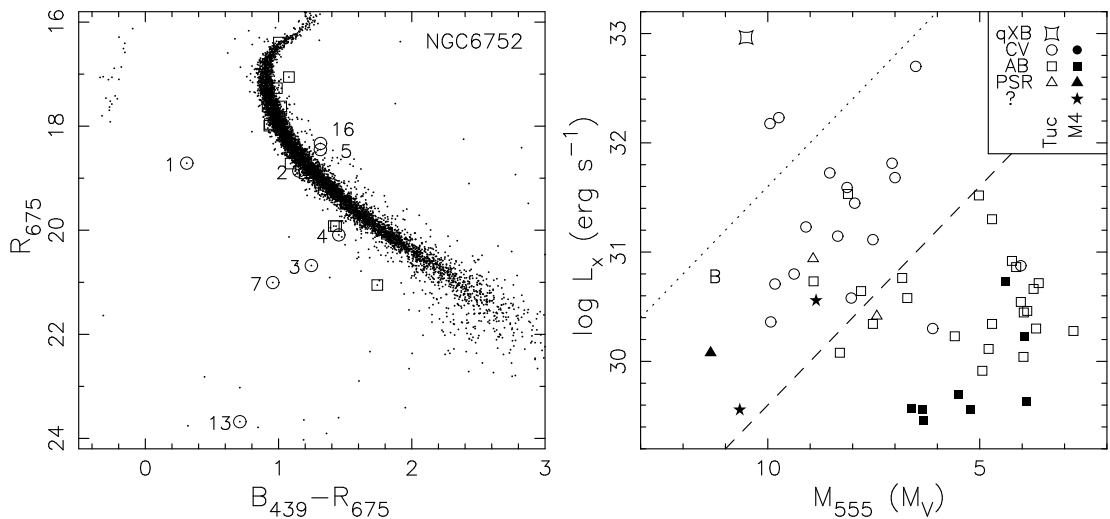} {\hfil}}

\caption{\it Left: Colour -- magnitude diagram for NGC6752. Optical
counterparts for X-ray sources are indicated $\odot$ and labelled with
the Chandra source number; other stars within the X-ray error circles, but
presumably not the counterparts are indicated $\Box$. From a re-analysis
by Bassa (private communication) after Pooley et al.\ (2002a).
Right: The X-ray luminosity as a function of absolute visual magnitude
for various source types in 47\,Tuc (open symbols) and M\,4 (filled
symbols). The dashed line roughly separates the cataclysmic variables (CV)
from the active binaries (AB). Other source types are quiescent low-mass
X-ray binaries (qXB), and companions to millisecond pulsars (PSR).
Sources which may be either a cataclysmic variable or an
active binary are indicated with a star. A 'B' indicates the position of 
an X-ray source (X2) in M\,4, which may be a background quasar.
The dotted line has $L_x$ hundred times the dashed line. Data from Edmonds 
et al.\ (2003), Grindlay et al.\ (2001) and Bassa et al.\ (2004).
\label{verbfe}
}
\end{figure}
\nocite{eghg03}\nocite{ghem01}\nocite{bph+04}

Once we have optical counterparts, we can complete the {\bf classification} 
of the sources. From the X-ray data alone we can classify the
quiescent low-mass X-ray binaries (from their spectrum and luminosity,
see Fig.\,\ref{verbfa}) and the pulsars (from positional coincidence).
In the optical colour-magnitude diagram, cataclysmic variables
are bluer than the main sequence, and active binaries are more
luminous than the main-sequence (Fig.\,\ref{verbfe}).
Cataclysmic variables have a higher X-ray to optical flux ratio
than magnetic binaries.
In a diagram of X-ray luminosity versus absolute visual magnitude, the line 
$$\log L_{0.5-2.5\,keV} {\rm (erg/s)} =-0.4M_V+33.6$$ 
roughly separates cataclysmic variables from active binaries
(Fig.\,\ref{verbfe}). 
The variability of cataclysmic variables
can cause confusion, if data are combined that were not taken
simultaneously. Also, background quasars may maskerade as cataclysmic
variables in the $L_x$-$M_V$ diagram, as may be the case for X2 in M\,4,
tentatively classified as a cataclysmic variable by Bassa et al.\ (2004),
but identical to the blue source, probably a quasar, discovered by
Bedin et al.\ (2003; Bedin \&\ King, private communication). This source is
indicated 'B' in Fig.\,\ref{verbfe}.
The dotted line, a factor 100 in X-ray luminosity above the dashed line,
roughly gives an upper limit to the X-ray luminosities of cataclysmic
variables in the galactic disk (based on ROSAT data, Verbunt et al.\ 1995).
Remarkably, two cataclysmic variables in 47\,Tuc are well above this line
(Grindlay et al.\ 2001).
\nocite{bpka03}\nocite{vbhj95}

\begin{figure}
\centerline{\psfig{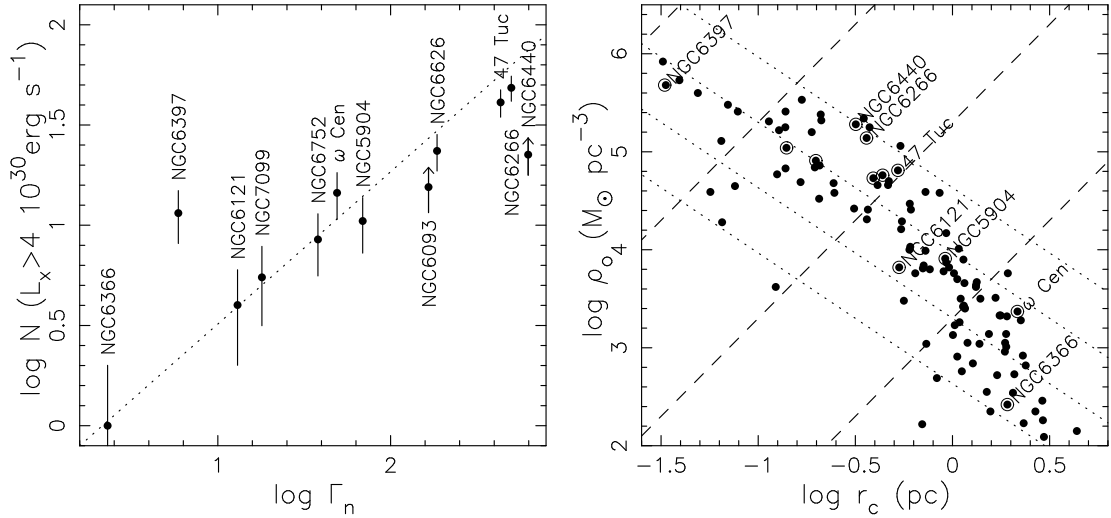} {\hfil}}

\caption{\it Left: The number of X-ray sources detected with Chandra
in globular clusters above a threshold limit, as a function of a
normalized collision number $\Gamma_n$.  An arrow indicates a lower
limit. The dashed line indicates the best fit:
$N\propto{\Gamma_n}^{0.74}$. After Pooley et al.\ (2003). Right:
central density of globular clusters as a function of core radius,
with lines of constant collision number
$\Gamma\propto{\rho_o}^{1.5}{r_c}^2$ (dotted, steps of factor 10) and
of constant destruction number $\gamma\propto{\rho_o}^{0.5}/r_c$
(dashed, steps of factor 10).  Clusters from the left panel are
encircled, and (some) indicated with their name. After Verbunt (2003).
\label{verbff}
}
\end{figure}
\nocite{pla+03}\nocite{ver03}

\section{Some considerations on formation rates}

If (quiescent) low-mass X-ray binaries and cataclysmic variables are
formed in globular clusters mainly through stellar encounters, their numbers
should scale with the collision number $\Gamma$ of the cluster. $\Gamma$
scales approximately with the central density $\rho_o$ and core radius
$r_c$ of the cluster as $\Gamma\propto{\rho_o}^{1.5}{r_c}^2$ (e.g.\ Verbunt 
2003). Pooley et al.\ (2003) show that this predictions holds rather well
for the X-ray sources with $L_x>4\times 10^{30}$\,erg\,s$^{-1}$ 
(in the 0.5-2.5\,keV band, Fig.\,\ref{verbff}). More precisely, with
$N\propto \Gamma^\alpha$, they find $\alpha=0.74\pm0.36$, allowing
a direct proportionality, but suggesting that $N$ increases slower than
$\Gamma$. Since most of these sources are actually cataclysmic
variables (see e.g.\ Fig.\,\ref{verbfe}), this indicates that most cataclysmic
variables in globular clusters are {\it not} evolved from primordial binaries,
but rather formed via stellar encounters.
NGC\,6397 is an interesting cluster in this respect, as it has a rather
larger number than predicted from its $\Gamma$ of binaries with neutron stars
(one quiescent low-mass X-ray binary, one radio pulsar) and
cataclysmic variables. Pooley et al.\ (2003) suggest that the cluster
was more massive in the past, and lost single stars from its outskirts
in passages close to the galactic center; binaries escape this stripping
process as they reside in the cluster core. The problem with this
explanation is that NGC\,6397 shows no excess of active binaries in it
core. Perhaps the explanation of this is that the active binaries are
destroyed in close encounters. The rate at which a single binary
undergoes an encounter is roughly $\gamma\propto {\rho_o}^{0.5}/r_c$
(Verbunt 2003), and NGC\,6397 has the highest $\gamma$ of the globular
clusters studied so far (Fig.\,\ref{verbff}). Since active binaries
tend to be wider than cataclysmic variables, they would be more
in danger of a destructive encounter.

A wonderful outlook on future work is provided by the study of bright
X-ray sources in the globular clusters of M\,87: Jord\'an et al.\ (2004)
show that the probability for a cluster to house a bright X-ray source
scales slower with $\rho_o$ than $\Gamma$. They argue that this may be
due to destruction of sources in the highest-density clusters.
\nocite{jcf+04}




\bibliographystyle{aipprocl} 


\end{document}